\documentclass[floatfix,preprint,pre,showpacs]{revtex4}
\usepackage{graphicx}
\usepackage{subfigure}
\usepackage{amssymb,amsmath}
\newcommand{\coloneq}{\; \colon \mspace{-12.0mu} =}

\usepackage[usenames]{color}

\begin{document}

\title{Numerical Results for the Blue Phases}
\author{G. P. Alexander${}^{a,b}$} 
\author{J. M. Yeomans${}^{a}$}
\affiliation{${}^{a}$The Rudolf Peierls Centre for Theoretical Physics, Oxford University, 1 Keble Road, Oxford OX1 3NP, U.K.}
\affiliation{${}^{b}$Department of Physics \& Astronomy, University of Pennsylvania, 209 South 33rd Street, Philadelphia PA 19104, U.S.A.}
\date{\today}


\begin{abstract}
We review recent numerical work investigating the equilibrium phase diagram, and the dynamics, of the cholesteric blue phases. In equilibrium numerical results confirm the predictions of the classic analytical theories, and extend them to incorporate different values of the elastic constants, or the effects of an applied electric field. There is a striking increase in the stability of blue phase I in systems where the cholesteric undergoes helical sense inversion, and the anomalous electrostriction observed in this phase is reproduced. Solving the equations of motion allows us to present results for the phase transition kinetics of blue phase I under dielectric or flexoelectric coupling to an applied electric field. We also present simulations of the blue phases in a flow field, showing how the disclination network acts to oppose the flow. The results are based on the Landau-de Gennes exapnsion of the liquid crystal free energy: that such a simple and elegant theory can predict such complex and subtle physical behaviour is remarkable. 
\end{abstract}
\maketitle



\section{Introduction}

One of the spectacular successes of de Gennes' Landau theory of liquid crystals is that it captures, without need for extension or modification, many of the remarkable and subtle phase transitions and properties of liquid crystals. Blue phases provide an especially vivid example of this. The general features and properties of these unique mesophases are both `beautiful and mysterious'~\cite{degennes93} and yet they can be understood qualitatively and quantitatively using de Gennes' ${\bf Q}$-tensor theory of liquid crystal phase transitions. 

Blue phases are found in highly chiral liquid crystals between the high temperature isotropic fluid and the lower temperature cholesteric phases~\cite{reinitzer88,degennes93,blumel85,yang87,crooker89,belyakov85,wright89}. They are remarkable mesophases, exhibiting a brightly coloured texture of individual, micron sized platelets. The bright colour indicates selective reflection due to a periodic structure, much like in an ordinary crystal, but with a much larger characteristic length scale, and indeed the reflection spectra show Bragg peaks that can be indexed by cubic space groups with lattice constants of several hundred nanometers. Furthermore, individual platelets of monodomain crystals themselves show distinctive facetting corresponding to Miller planes of the lattice structure and with the faces growing in a sequence of steps~\cite{pieranski85b}. Yet, blue phases are not crystals in the traditional sense: they have {\it no} long range positional order and are full three-dimensional fluids. The crystalline order is in the orientational degrees of freedom of the liquid crystal. 
  
The key to understanding the properties of the blue phases was in realising that the locally preferred order in the chiral liquid crystal is one of {\em double twist}, with the molecules adopting helical ordering along two, perpendicular axes, as opposed to the usual single twist of an ordinary cholesteric helix~\cite{meiboom81,meiboom83,hornreich81,hornreich82}. However, it is not possible to construct a global state with helical ordering in two directions, without introducing disclination lines into the structure. Therefore the blue phases consist of local, cylindrical, regions of double twist separated by a regular disclination network.

\begin{figure}
\centering
\includegraphics[width=80mm]{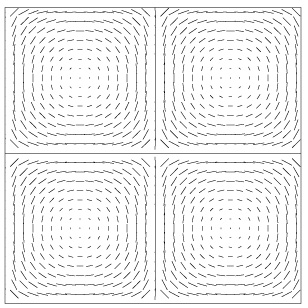} 
\caption{Director field in a hypothetical, two-dimensional, blue phase showing how local regions of double twist can be pieced together with a square array of topological defects. (Blue phases with two-dimensional symmetry and translational invariance in the third dimension have been observed experimentally in an electric field \cite{pieranski85}, but these have hexagonal symmetry.)}
\label{square_blue_phase}
\end{figure}

\begin{figure}
\centering
\includegraphics[width=.95\textwidth]{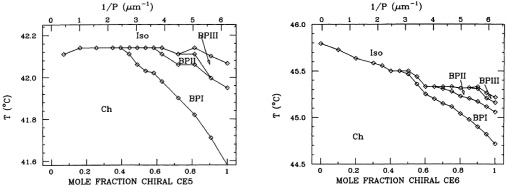}
\caption{Experimental phase diagrams of the cholesteric blue phases for two different chiral compounds, reproduced from Reference~\cite{yang87}. Three distinct blue phases are found, in the order blue phase I, blue phase II, blue phase III upon increasing the amount of chiral dopant.}
\label{phase_diagram}
\end{figure}

\begin{figure}
\centering
\includegraphics[width=.95\textwidth]{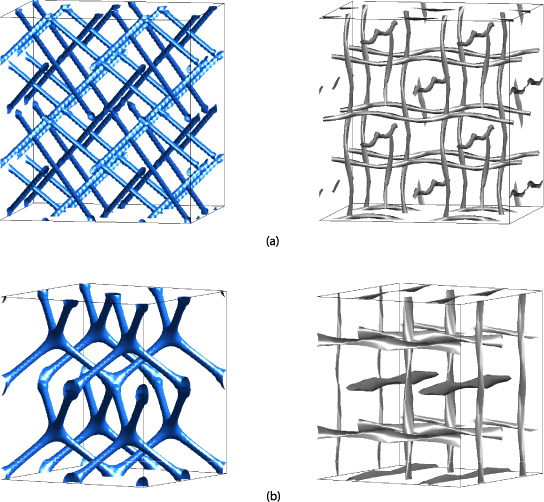}
\caption{Network of disclinations and arrangement of double twist cylinders in blue phase I (a) and blue phase II (b). The disclinations are shown on the left in blue and the double twist cylinders on the right in grey. In all figures $2^3$ unit cells have been shown to more clearly illustrate the structure.}
\label{structures}
\end{figure}

A tutorial, two-dimensional example of a blue phase texture is shown in Fig.~\ref{square_blue_phase}. This demonstrates how local regions of double twist can be pieced together only by introducing a square array of topological defects. In physical, three-dimensional, crystals the disclination structures are more complex; three distinct blue phase textures have been identified upon cooling from the isotropic liquid. Typical experimental phase diagrams are shown in Fig.~\ref{phase_diagram}. Blue phases I and II exhibit textures with cubic symmetry corresponding to the space groups $O^{8-}$ and $O^2$ respectively~\cite{wright89}. These are illustrated in Fig.~\ref{structures} which shows both the disclination networks and the topology of the lattice of double twist cylinders in the two phases. Blue phase III is less well understood; it has an amorphous structure with the same symmetry as the isotropic fluid~\cite{lubensky96,anisimov98}.

The unique combination of crystalline order with lattice constants comparable to the wavelength of visible light and full three dimensional fluidity make the blue phases ideal for technological uses such as fast light modulators, photonic crystals or tunable lasers~\cite{cao02,hisakado05,yokoyama06,kitzerow06}. The principal obstruction to their application was the very limited temperature range, $\sim 1K$, over which the phases were stable. However, recently, the stability range has been extended to as much as $60K$, including room temperature, by the addition of bimesogenic molecules or photo-crosslinking of polymers~\cite{kikuchi02,coles05}. This has opened the way to device applications and, in May 2008, Samsung Electronics unveiled a prototype `blue phase mode LCD'~\cite{wikiBPLCD08,kikuchi07b}.

The theoretical description of the general features and properties of the blue phases was developed in the 1980s in a series of important papers, summarised in the review \cite{wright89}. Two approaches were followed, a low chirality theory, based on the Frank director field description~\cite{meiboom81,meiboom83} and a high chirality theory using the Landau-de Gennes $\bf Q$-tensor~\cite{grebel83,grebel84}. The theoretical approaches were able to correctly predict the symmetries of the blue phases and to give a good account of the phase diagram, although with some discrepancies. For example, additional textures with space groups $O^{8+}$ and $O^5$ were found to be stable within the parameter range corresponding to the experiments, but were not observed experimentally.

The renewed experimental interest in the blue phases provided the motivation to revisit them theoretically. Since the original analytic investigations, computers have become a powerful tool for studying complex fluids, and it is now feasible not only to revisit and extend the calculations of the blue phase phase diagrams~\cite{dupuis05a,gareth06}, but also to obtain results on blue phase kinetics and hydrodynamics~\cite{dupuis05b,gareth08}. The aim of this article is to review recent research showing how numerical simulations of the blue phases are allowing us to gain further insights into their behaviour.

In Section II we describe the Landau-de Gennes equations that have proved so successful in modelling the blue phases and outline how they can be solved numerically. In particular we discuss how the size of the unit cell must be allowed to vary to correctly minimise the free energy. Then, in Section III, we revisit the classic calculations of the thermodynamic phase diagram showing the changes that result when the approximations necessary to make the analytic calculations feasible can be circumvented by a numerical solution. We also show that including additional terms in the Landau-de Gennes free energy can significantly change the phase diagram. Most dramatic among the results is the increase in the stability of blue phase I obtained in systems where the cholesteric undergoes helical sense inversion. 

Many of the potential technological applications of the blue phases rely on the response of the material to an electric field. In the blue phases the effects of an electric field are particularly rich, including continuous distortions in the size and shape of the unit cell~\cite{stegemeyer84,porsch86,porsch89} and a series of field-induced transitions to new blue phase structures, not stable in zero field~\cite{pieranski85,cladis86,pieranski87,chen87}.  In Section IIIC we show how the Landau-de Gennes expansion correctly predicts electrostriction and, in particular, the anomalous electrostriction of blue phase I.

We next, in Section IV, summarise the equations of motion which describe blue phase hydrodynamics. These can be used to investigate the kinetics of transitions between blue phases, and their response to an applied electric field or to an imposed flow. Section V is devoted to examples demonstrating the application of the dynamical equations. We describe an investigation into the viscoelastic properties of the blue phases as they are subject to a Poiseuille flow field. We then describe the kinetics of the phase transitions induced in blue phase I as it is coupled to an electric field, either via a dielectric, or a flexoelectric, term in the free energy. This allows us to propose a candidate structure for blue phase X.

\section{Minimising the Landau-de Gennes free energy}

\subsection{Landau expansion in terms of the $\bf Q$-tensor}

One of de Gennes' vital contributions to the theory of liquid crystals was to identify the $\bf Q$ tensor as a suitable order parameter~\cite{degennes93}. A Landau expansion of the free energy of liquid crystals in terms of $\bf Q$ has proved useful as a starting point for interpreting liquid crystal phase diagrams~\cite{degennes93,wright89,gareth06,grebel83,brazovskii75}
\begin{equation}
\begin{split}
F = \tfrac{1}{V} \int_{\Omega} \text{d}^3r \Biggl\{ & \tfrac{A_0(1 - \gamma /3)}{2} \, \text{tr} \, \bigl( {\bf Q}^2 \bigr) - \tfrac{A_0 \gamma}{3} \, \text{tr} \, \bigl( {\bf Q}^3 \bigr) + \tfrac{A_0 \gamma}{4} \Bigl( \text{tr} \, \bigl( {\bf Q}^2 \bigr) \Bigr)^2 \\
& \; + \tfrac{L_{21}}{2} \Bigl( \nabla \times {\bf Q} + 2q_0 {\bf Q} \Bigr)^2 + \tfrac{L_{22}}{2} \Bigl( \nabla \cdot {\bf Q} \Bigr)^2 + q_0 L_{31} {\bf Q}^2 \cdot \nabla \times {\bf Q} \\
& \; + \tfrac{L_{34}}{2} Q_{\alpha \beta} \nabla_{\alpha} Q_{\gamma \delta} \nabla_{\beta} Q_{\gamma \delta} + \tfrac{L_{38}}{2} Q_{\alpha \beta} \nabla_{\gamma} Q_{\alpha \delta} \nabla_{\gamma} Q_{\beta \delta} \Biggr\} .
\end{split}
\label{free}
\end{equation} 
Here $A_0$ is a constant with the dimensions of an energy density, $\gamma$ plays the role of an effective temperature for thermotropic liquid crystals, $q_0$ defines the helical pitch and the $L_{ij}$ are elastic constants. The expansion, taken to second order in the derivatives of $\bf Q$, can only account for two independent Frank elastic constants and imposes that the magnitude of splay and bend are equal~\cite{wright89}. In order to remove this constraint, and to allow for a temperature dependent helical pitch, it is necessary to consider, at least, terms cubic in $\bf Q$ and quadratic order in gradients~\cite{berreman84,longa87,longa89}. There are eight such terms and it is unrealistic to consider them all. We include one chiral $L_{31}$ and two achiral terms $L_{34}$ and $L_{38}$.  The $L_{34}$ term was chosen as it contributes equally to all three elastic constants and the $L_{38}$ term as it gives the largest distinction between splay and bend and because it has the largest contribution to the energetics of an isolated double twist cylinder. (However, the choice is arbitrary and other coefficients may be non-zero in any given material, and may have a considerable effect on the phase diagram.)

Rewriting the free energy in terms of dimensionless variables demonstrates that it depends only on the two dimensionless parameters~\cite{gareth06}
\begin{equation}
\tau \coloneq \frac{9(3 - \gamma)}{\gamma} , \hspace{20mm} \kappa^2 \coloneq \frac{108q_0^2L_{21}}{A_0\gamma} , 
\label{eq:deftau}
\end{equation}
known as the {\it reduced temperature} and the {\it chirality} respectively, together with ratios of the elastic constants. In line with previous work, phase diagrams will be presented in terms of these parameters.

\subsection{Numerical minimisation of the free energy}

The free energy \eqref{free} is minimised by relaxing the $\bf Q$-tensor according to a Ginzburg-Landau equation 
\begin{equation}
\partial_t {\bf Q} \; = \; \Gamma \, \biggl( \frac{-\delta F}{\delta {\bf Q}} + \tfrac{1}{3}\, \text{tr} \biggl( \frac{\delta F}{\delta {\bf Q}} \biggr) \, {\bf I} \biggr) . 
\label{relax}
\end{equation}
This equation can be solved using many different numerical approaches; the results presented here were obtained using a lattice Boltzmann algorithm~\cite{denniston04,julia06}.

To study the different blue phases it is necessary to implement appropriate initial conditions for the simulation. The ${\bf Q}$-tensor is initialised using analytic expressions appropriate to the high chirality limit ($\kappa \rightarrow \infty$), which act to define the symmetry of the chosen phase. For blue phase I~\cite{grebel83,wright89,dupuis05a} we use 
\begin{equation}
\begin{split}
Q_{xx} & \sim - \sin (ky/\sqrt{2}) \cos (kx/\sqrt{2}) - \sin (kx/\sqrt{2}) \cos (kz/\sqrt{2}) \\
& \quad + 2 \sin (kz/\sqrt{2}) \cos (ky/\sqrt{2}) , \\
Q_{xy} & \sim -\sqrt{2} \sin (kx/\sqrt{2}) \sin (kz/\sqrt{2}) - \sqrt{2} \cos (ky/\sqrt{2}) \cos (kz/\sqrt{2}) \\
& \quad + \sin (kx/\sqrt{2}) \cos (ky/\sqrt{2}) , 
\end{split}
\label{eq:QO8-}
\end{equation} 
where $k = 2\sqrt{2}\pi /a$, with $a$ the lattice constant, and the other components are obtained by cyclic permutation. Similarly, for blue phase II the ${\bf Q}$-tensor is initialised as~\cite{grebel83,wright89,dupuis05a}  
\begin{equation}
\begin{split}
Q_{xx} & \sim \cos (kz) - \cos (ky) , \\
Q_{xy} & \sim \sin (kz) , 
\end{split}
\label{eq:QO2}
\end{equation} 
where $k = 2\pi /a$ and the other components are again obtained by cyclic permutation. 

Under numerical evolution using Eq.~\eqref{relax} the system relaxes to the structure of the same symmetry that locally minimises the free energy. We are therefore able to obtain, for any value of the parameters, local minima of the free energy corresponding to each of the cholesteric and blue phases. The global free energy minimum was taken to be the smallest of these calculated local minima. 

To achieve a full minimisation of the free energy it is necessary to set the correct unit cell size in the simulation. This is not known {\it a priori}, but rather depends on the magnitude of the order parameter, a quantity that is only determined by the numerical minimisation itself. Therefore we must introduce a means of determining, and setting, the unit cell size as the ${\bf Q}$-tensor evolves during the simulation. It is possible to account for a change in unit cell size by rescaling the gradient contributions to the free energy and molecular field. This is accomplished in practice by changing the elastic constants to 
\begin{equation}
\begin{split}
q_0 & = q_0^{\text{init}}/r , \\
L_{ij} & = L_{ij}^{\text{init}} \times r^2 ,
\end{split}
\label{eq:rescalingtrick}
\end{equation}
where a superscript `init' denotes the initial value of a simulation parameter and $r$ is the appropriate rescaling factor, which is identical to the `redshift' described in ~\cite{grebel83,grebel84}.

To calculate the optimal size of the unit cell we note that, since the free energy is quadratic in gradients, it may be written formally in ${\bf k}$-space as
\begin{equation}
f = ak^2 + bk + c ,
\label{eq:fredshifttrick}
\end{equation}
where the coefficients $a,b$ and $c$ depend on the ${\bf Q}$-tensor, but not on ${\bf k}$. The optimum wavevector is given by $k = -b/2a$, and since the coefficients $a$ and $b$ are determined by the simulation it is straightforward to use these values to determine the exact value for the size of the unit cell at every timestep, thereby obtaining a full minimisation of the free energy.

\section{The equilibrium phase diagram}

\subsection{Revisiting the analytic calculations}

The phase diagram for chiral liquid crystals obtained for a selection of parameter values using the Landau-de Gennes free energy, Eq.~\eqref{free}, is shown in Fig.~\ref{fig:BPphasediagram}. Fig.~\ref{fig:BPphasediagram}(a), adapted from \cite{grebel84}, shows the phase diagram calculated analytically in the high chirality limit, in the one elastic constant approximation, and Fig.~\ref{fig:BPphasediagram}(b) compares numerical results for the same parameters in the free energy. The differences between the analytic and numerical results show that including higher order harmonics in the minimisation of the free energy is significant, as might be expected for phases with small free energy differences. 
In the numerical (exact) minimisation $O^{8-}$ is stable over a larger range of parameters and at lower chirality values, in better agreement with the experimental phase diagram, Fig~\ref{phase_diagram}. Moreover, the regions of stability of the $O^5$ and $O^{8+}$ textures found in the early studies~\cite{grebel83,grebel84} are shifted to unphysically high values of the chirality~\cite{dupuis05a}. This result is again consistent with experiment, where blue phase structures of these symmetries have not been observed.

\subsection{Varying the elastic constants}

\begin{figure}
\begin{center}
\includegraphics[width=120mm]{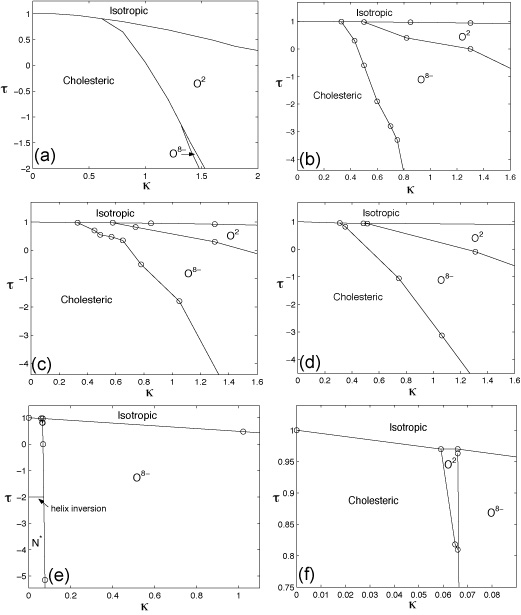}
\caption{Phase diagram of cholesteric liquid crystals within the Landau-de Gennes theory. $\tau$, the reduced temperature, and $\kappa$, the chirality, are defined by Eq.~\eqref{eq:deftau}. Top row: phase diagrams in the one elastic constant approximation determined (a) analytically using truncated Fourier series, and (b) numerically. Middle row: numerical phase diagrams with unequal elastic constants: (c) $K_1=K_2=0.5 K_3$, (d) $K_1=K_3=1.5 K_2$. Bottom row: (e) the numerical phase diagram obtained with the chiral invariant, $L_{31}$, added to the free energy. The magnitude of this term was chosen so as to produce helical sense inversion in the cholesteric phase at a temperature not far below the isotropic transition temperature. (f) an enlargement of the region near the isotropic transition temperature. Note the reversal in the order of appearance of blue phase I and blue phase II as a function of chirality. (adapted from \cite{gareth06})}
\label{fig:BPphasediagram}
\end{center}
\end{figure}

We now compare the phase diagrams obtained as the ratios of the elastic constants are varied. To investigate the effect of the bend elastic constant we chose parameter values $L_{21}=L_{22}=L_{34}=0.02, \, L_{38}=0$ which corresponds to a ratio of splay to bend of about 0.5, while splay and twist remain degenerate. The resulting phase diagram is shown in Fig.~\ref{fig:BPphasediagram}(c).  Comparing to the case of equal elastic constants the stability of blue phase I is seen to decrease quite significantly relative to the cholesteric phase while at the same time there is a small increase in stability over blue phase II. There is only a minor shift in the cholesteric-blue phase I phase boundary at the transition temperature, however, as the temperature decreases the shift becomes larger. 

The value of the twist elastic constant is controlled by the Landau-de Gennes parameter $L_{22}$. In most liquid crystals the twist elastic constant is smaller than either splay or bend. In order to match this, we constructed the phase diagram for parameter values $L_{21}=0.02, \, L_{22}=0.04, \, L_{34}=L_{38}=0$, which is shown in Fig.~\ref{fig:BPphasediagram}(d). This choice of parameters resulted in a ratio of splay to twist of about 1.5, while splay and bend remained degenerate. Again we observe that the stability of blue phase I is reduced relative to the cholesteric phase by an amount similar to that seen by varying the bend elastic constant. 

Finally, we consider the effect of the chiral cubic invariant on the blue phases. We chose parameter values of $L_{21}=L_{22}=L_{34}=0.02, \, L_{38}=0.04, \, L_{31}=-0.18$, which gives a ratio of bend to splay of about $1.6$. Moreover, for these coefficients the cholesteric undergoes helical sense inversion~\cite{huff00,slaney92} at a reduced temperature of about $\tau \approx -2$. The phase diagram is shown in Fig.~\ref{fig:BPphasediagram}(e). What is remarkable is the dramatic increase in stability of blue phase I relative to the cholesteric phase. The region of stability has been increased down to chiralities as low as $\kappa = 0.07$ and at such low chiralities the phase boundary is essentially independent of $\kappa$ for all $\tau$. In addition, there is a very small region of stability for blue phase II located close to the isotropic transition (Fig.~\ref{fig:BPphasediagram}(f)). Since blue phase I is now stable over a much larger temperature range it displays a significant variation in unit cell size as the temperature is lowered, with the lattice parameter more than doubling between $\tau=1$ and $\tau=-5$.

\subsection{Electrostriction}

The response of the cubic blue phases to the application of an external field has been a topic of interest since the mid 1980s and continues to be so because of the importance to potential blue phase based devices~\cite{kitzerow91,kitzerow06,kikuchi07b}. The principal features include electrostriction, a continuous distortion of the shape and size of the unit cell with increasing field, and a series of field induced textural transitions. The electrostriction involves a shift of the back-scattered Bragg peak of 5--10\% and is quadratic in the field strength~\cite{stegemeyer84,porsch86,porsch89}. The direction of the shift changes sign with the sign of the dielectric anisotropy, but blue phase I also displays an unusual response referred to as anomalous electrostriction, where an expansion along the field direction is seen when the field is applied parallel to the $[011]$ direction, but a contraction for fields parallel to $[001]$. At larger field strengths new blue phases appear. Three distinct field induced textures have been identified, possessing tetragonal, screw hexagonal and two-dimensional hexagonal symmetry with increasing field strength~\cite{pieranski85,cladis86,pieranski87,chen87}. 

Much of the electric field behaviour has been understood theoretically via extensions of the Landau-de Gennes theory, including the qualitative features of the electrostriction and the field induced textural transitions~\cite{hornreich85,lubin87,stark91,hornreich90,longa96,zelazna98}. However, the approximations inherent in the analytic calculations limited the quantitative comparison that was possible and a number of features, including the anomalous electrostriction of blue phase I and the tetragonal field induced texture blue phase X, could not be accounted for~\cite{zelazna98,hornreich90}.  

The main additional difficulty in minimising the free energy of the blue phases in the presence of an electric field~\cite{gareth08} is in accounting for electrostriction, as there is a change not only in the size, but also in the shape, of the unit cell as the field is applied. This distortion can be accounted for in two steps: first the shape of the simulation unit cell is fixed and the corresponding size, which minimises the free energy, is determined. This is then repeated for a set of varying shapes of the unit cell, e.g. from cubic to tetragonal. Hence the free energy is determined for several values of the ratio $L_z/L_x$ parameterising the cubic-tetragonal distortion and fitted to a quadratic. The actual distortion is then given by the miminum of this fit. 

\begin{figure}
\begin{center}
\includegraphics[width=.65\textwidth]{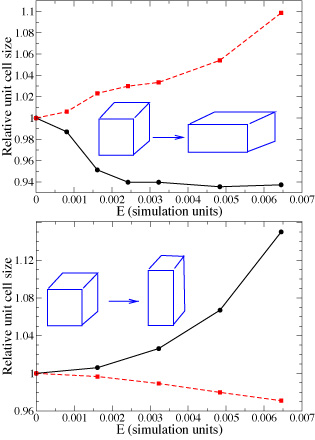}
\caption{Distortion of the unit cell of blue phase I (top) and blue phase II (bottom) in an electric field applied along [001]. The lattice parameters parallel ($\circ$, solid line) and perpendicular ($\square$, dashed line) to the field are shown relative to their zero field value. Note that blue phase I contracts parallel to the field, whereas blue phase II expands parallel to the field. (adapted from \cite{gareth08})}
\label{fig:electrostriction}
\end{center}
\end{figure}

The field dependence of the blue phase lattice parameters is shown in Fig.~\ref{fig:electrostriction} for an applied electric field along the $[001]$ direction. Note, in particular, that the unit cell expands along the field direction in blue phase II, but contracts in blue phase I, in agreement with experiment~\cite{kitzerow91}. When the field is instead applied along to the $[011]$ direction both blue phases undergo an expansion parallel to the field, a precursor to the transition to blue phase X that is observed at larger field strengths. It is very pleasing that a numerical approach can predict anomalous electrostriction in blue phase I as the effect is lost in the truncations needed in analytic calculations~\cite{zelazna98}. Mapping from physical to simulation units gives a magnitude of the elctrostriction in the range $10^{-2} - 10^{-1} \mu m^2\, V^{-2}$ for both blue phases, again in good agreement with experiments~\cite{kitzerow91}.

\section{Hydrodynamic equations}

The Ginzburg-Landau equation, Eq.~\eqref{relax}, describes the relaxation of a liquid crystal to the minimum free energy, but does not describe physical dynamics in situations where flow is important. Several authors have recently used numerical approaches to solve the full hydrodynamic equations of motion for liquid crystals in the nematic and cholesteric phases. For example, it has been possible to simulate defect hydrodynamics~\cite{toth02,toth03,svensek02,svensek03}, phase ordering~\cite{denniston00,denniston01c,nidhal06}, the kinetics of transitions between different liquid crystal phases~\cite{fukuda98,gareth08}, cholesteric rheology~\cite{davide04a,davide04b,orlandini05,davide05b,davide06,orlandini07} and the effect of flow on device switching~\cite{qian97,qian01,denniston01a,toth02b,denniston02,davide03,davide05a,stromer06,james08}. Because of their disclination structure the kinetics and the rheology of the blue phases is an exciting, but demanding, numerical problem which requires intensive numerical resources. However, such simulations are rapidly becoming feasible: we summarise the equations of motion and some of the results obtained so far, and then discuss possible directions for future work. 

The hydrodynamic equations of motion of liquid crystals are complex, both because of the anisotropy of the molecules and because of the coupling between the order parameter field and the flow field. In general, flow leads to a rotation of the local orientation, which in turn influences the flow. Similarly, if a disturbance is initiated in the director, its reorientation is generally accompanied by fluid motion, an effect sometimes referred to as {\it backflow}. If a system is close to a phase transition, or contains disclinations, variations in the magnitude of the order parameter can be significant and therefore a hydrodynamic description based on the ${\bf Q}$-tensor is needed~\cite{beris94,grmela97,ottinger97,ottinger05,qian98}.  

The order parameter evolves towards the minimum of the free energy, but with a convective time derivative to account for the advection with the fluid  
\begin{equation}
D_t {\bf Q} \; = \; \Gamma \, \biggl( \frac{-\delta F}{\delta {\bf Q}} + \tfrac{1}{3}\, \text{tr} \biggl( \frac{\delta F}{\delta {\bf Q}} \biggr) \, {\bf I} \biggr)  .
\label{eq:Qevol}
\end{equation} 
The term in brackets on the right hand side is called the molecular field, ${\bf H}$, and $\Gamma$ is a collective rotational diffusion constant. The material derivative for rod-like molecules is given by~\cite{beris94} 
\begin{equation}
\begin{split}
D_t {\bf Q} & = \bigl( \partial_t + {\bf u} \cdot \nabla \bigr) \, {\bf Q} - \bigl( \xi {\bf D} + \boldsymbol{\Omega} \bigr) \, \bigl( {\bf Q} + \tfrac{1}{3} {\bf I} \bigr) - \bigl( {\bf Q} + \tfrac{1}{3} {\bf I} \bigr) \, \bigl( \xi {\bf D} - \boldsymbol{\Omega} \bigr) \\
& \quad + 2\xi \, \bigl( {\bf Q} + \tfrac{1}{3} {\bf I} \bigr) \, \text{tr} \, \bigl( {\bf QW} \bigr) \; ,
\end{split}
\label{eq:materialderivativeQ}
\end{equation}
where ${\bf D} = ({\bf W} + {\bf W}^{T})/2$ and $\boldsymbol{\Omega} = ({\bf W} - {\bf W}^{T})/2$ are the symmetric and antisymmetric parts, respectively, of the velocity gradient tensor $W_{\alpha \beta} = \nabla_{\beta} u_{\alpha}$. The constant $\xi$ depends on the molecular details of a given liquid crystal. 
 
The fluid velocity field is taken to obey the continuity equation and a Navier-Stokes equation with a stress tensor generalised to describe liquid crystal hydrodynamics~\cite{beris94}  
\begin{gather}
\partial_t \varrho + \nabla \cdot \bigl( \varrho {\bf u} \bigr) = 0 \; , \label{eq:continuity} \\
\varrho \bigl( \partial_t {\bf u} + {\bf u} \cdot \nabla {\bf u} \bigr) = - \nabla p + \nabla \cdot \boldsymbol{\sigma} \; , \label{eq:NavierStokes} \\
\begin{split}
\sigma_{\alpha \beta} & = \mu \bigl( \nabla_{\alpha} u_{\beta} + \nabla_{\beta} u_{\alpha} \bigr) + 2 \xi \Bigl( Q_{\alpha \beta} + \tfrac{1}{3} \delta_{\alpha \beta} \Bigr) \, Q_{\gamma \delta} H_{\gamma \delta} \\
& \quad - \xi H_{\alpha \gamma} \Bigl( Q_{\gamma \beta} + \tfrac{1}{3} \delta_{\gamma \beta} \Bigr) - \xi \Bigl( Q_{\alpha \gamma} + \tfrac{1}{3} \delta_{\alpha \gamma} \Bigr) H_{\gamma \beta} \\
& \quad + Q_{\alpha \gamma} H_{\gamma \beta} - H_{\alpha \gamma} Q_{\gamma \beta} - \nabla_{\alpha} Q_{\gamma \delta} \, \frac{\delta F}{\delta \nabla_{\beta} Q_{\gamma \delta}} \; .
\end{split}
\label{eq:BEstress}
\end{gather}

Eqs.~\eqref{eq:Qevol},~\eqref{eq:continuity} and~\eqref{eq:NavierStokes} can be solved either using a lattice Boltzmann approach~\cite{denniston01d,denniston04}, or finite difference schemes~\cite{svensek02,james06,qian01} or, giving somewhat improved stability while retaining the advantages of lattice Boltzmann, a hybrid scheme where a lattice Boltzmann solution of the flow equation is coupled to a finite difference solution of the order parameter field~\cite{davide07}.

\section{Blue phase rheology}

\subsection{Response of the disclination lattice to a Poisueille flow}

\begin{figure}
\begin{center}
\includegraphics[width=120mm]{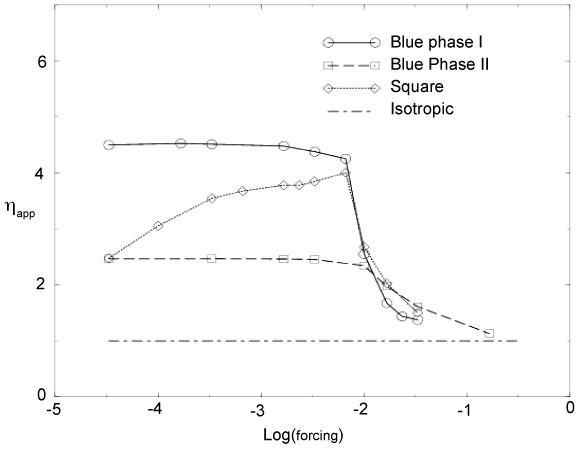}
\caption{Variation of the apparent viscosity of blue phase I, blue phase II, and a square array of double twist cylinders as a function of the applied forcing. The viscosity of the isotropic liquid crystal is also shown. Initially the blue phases show an enhanced viscosity as the disclination network opposes the flow. At higher forcing the disclination network breaks up and the viscosities of the blue phases tend to that of the isotropic liquid crystal. (adapted from \cite{dupuis05b})}
\label{fig:viscosity}
\end{center}
\end{figure}

\begin{figure}
\begin{center}
\includegraphics[width=120mm]{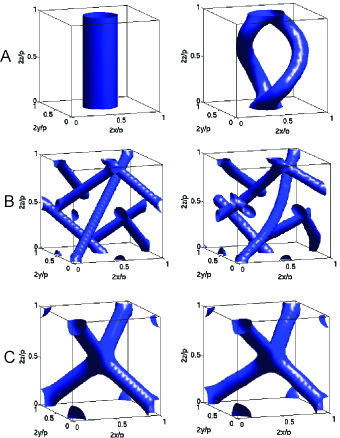}
\caption{Structure of the blue phases under Poiseuille flow. Rows A, B and C correspond to a square array of double twist cylinders, blue phase I, and blue phase II respectively. The first column shows the disclination network at zero forcing and the second a steady state of the network under flow. Note how, in row A, the strength 1 disclination opens to a strength $\frac{1}{2}$ ring. (adapted from \cite{dupuis05b})}
\label{fig:flow}
\end{center}
\end{figure}

To investigate the rheological response of the blue phases we placed a unit cell between fixed plates and imposed constant force on the fluid, together with no-slip boundary conditions on the velocity field at the plates. In a Newtonian fluid this geometry leads to a quadratic, Poiseuille flow profile. The choice of boundary conditions for the director field was to assume that the disclinations are fixed at the boundaries. Fig.~\ref{fig:viscosity} compares the apparent viscosity (obtained comparing to the Poiseuille velocity) in blue phases I and II, in a phase comprising a square array of disclinations with two-dimensional cross section corresponding to Fig.~\ref{square_blue_phase}, and in the isotropic phase. The corresponding disclination configurations, comparing zero and a finite velocity field, are shown in Fig.~\ref{fig:flow}.

For small forcing the blue phase viscosities increase by a factor $\sim 4$ over that of the isotropic fluid. This is because the disclination network acts to oppose the flow and dissipate energy. The blue phases reach a stationary state in which the disclination network is bent and twisted by the flow. The viscosities of blue phases I and II are approximately constant over a range of forcing, but the square lattice structure shows shear thickening. This is because each defect line, of topological strength 1, opens to a disclination ring, comprising disclinations of strength 1/2. The ring then twists and bends on itself as the flow increases, as shown in Fig.~\ref{fig:flow}.

As the forcing increases there is significant shear thinning in all three blue phase structures. This occurs because the disclination network is destroyed by the flow, and the viscosity drops to that of an isotropic liquid crystal.

These results indicate that blue phase rheology is extremely rich, and worthy of further study, both experimentally and numerically. There is a need for careful experiments, with good control over boundary conditions. On the numerical side, work is in progress to move towards larger numbers of unit cells in the simulations and to assess the effects of different flow geometries and boundary conditions~\cite{henrich09}.

\subsection{Blue phase X}

The ability to solve the dynamical equations of motion means that it is possible to follow the way in which disclinations rearrange during continuous textural transitions. Moreover, the simulations provide a way to identify possible candidate structures for blue phases which have not yet been identified.

\begin{figure}
\centering
\includegraphics[width=60mm]{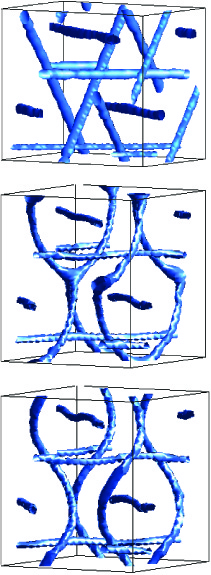} 
\caption{Simulation of the blue phase I--blue phase X transition. Top: blue phase I before the field is applied. Middle: at the transition from blue phase I to blue phase X. Bottom: the final configuration of disclination lines in blue phase X. In all cases the electric field is applied parallel to the $[011]$ direction, which is vertical in the figure. (adapted from~\cite{gareth08})}
\label{blue_phase_X}
\end{figure}

An example of this approach is simulations in which blue phase I is placed under an intermediate electric field applied parallel to the $[011]$ direction. This is the set-up corresponding to the blue phase I-blue phase X transition observed experimentally in the 1980s~\cite{cladis86,pieranski87}. The electrostriction distorts the shape of the unit cell until it becomes tetragonal, at which point there is a transition from the blue phase I texture to a new texture known as blue phase X. Starting from this geometry, and an unperturbed blue phase I, the evolution of the disclination network is simulated numerically. Initially, the disclinations in the network twist, they then merge to form a transiently branched structure, which finally reorganises into a new defect network, not stable at zero field as shown in Fig.~\ref{blue_phase_X}. This is a candidate structure for blue phase X, as it is a new network, found via a continuous reorganisation starting from blue phase I, and only stable in a field. The results predict that in the candidate blue phase X (i) the disclinations perpendicular to the field are largely unaffected, (ii) the network conforms to the space group $D_4^{10}$ identified in Reference~\cite{pieranski87}, and (iii) the double twist cylinders deform but do not break during the transition; this observation may be verified by experiments along the lines of those in References~\cite{kikuchi07,higashiguchi08}.

\subsection{Flexoelectric blue phases}

In 1969 Meyer introduced the concept of flexoelectric coupling to an external electric field and showed that this could lead to a one-dimensional splay-bend distortion of the nematic director field~\cite{meyer69}. These results have recently been extended to show that, near the isotropic-nematic transition and with sufficiently strong coupling, two-dimensional splay-bend structures with hexagonal symmetry can be stable~\cite{gareth07}. 

\begin{figure}
\centering
\includegraphics[width=160mm]{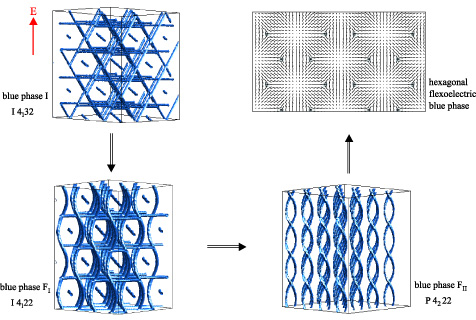} 
\caption{Textural transitions in blue phase I induced by flexoelectricity. The electric field is applied parallel to the $[011]$ direction (vertical in the figure) and produces a series of transitions first to two distinct textures possessing tetragonal symmetry, with space groups $I4_122$ and $P4_222$, and finally to a two dimensional hexagonal texture (viewed here along the field direction) at larger field strengths.}
\label{flexo_fig2}
\end{figure}

Fig.~\ref{flexo_fig2} shows the evolution of blue phase I when flexoelectric coupling to an applied field is increased quasistatically. As in the dielectric case, the electric field induces a twist in the disclination lines allowing them to transiently merge, thus fascilitating textural transitions. For the largest applied field strengths a transition is indeed observed to the two-dimensional hexagonal flexoelectric blue phase, as expected. However, at intermediate values, two further transitions precede it, each yielding distinct textures that are stable over a small range of field strengths. The first is to a centred tetragonal texture with space group $I4_122$. This has the same disclination network as the blue phase X structure found under dielectric switching and described above. At slightly higher field strengths there is a second transition to a distinct tetragonal texture with space group $P4_222$. In this texture, the disclinations are all parallel to the field, occuring in pairs that wrap around each other to form a double helix, and with the axes of the double helices themselves then arranged on a square lattice. Finally, as the flexoelectric coupling is inceased, the two members of a given double helix transiently merge and re-separate, allowing them to both straighten out and to adopt an hexagonal configuration in the plane perpendicular to the field.

\section{Discussion}

The Landau-de Gennes expansion of the free energy of liquid crystals has proved a vital tool in understanding their thermodynamics. The theory has proven successful even when very subtle energy-entropy balances, such as those that stabilise the blue phases, come into play. Beautiful early calculations, using approximate theories based on the Landau-de Gennes formalism that are analytically tractable, helped understand many of the features of the blue phases. Here we have shown that it is possible to make further progress by exploiting modern computational resources to minimise the free energy exactly. 
   
We have shown that significant quantitative differences in the phase diagram arise from retaining cubic order terms in the free energy expansion. In particular, choosing expansion coefficients appropriate to systems where the cholesteric undergoes helical sense inversion gives rise to a very significant increase in the stability of blue phase I. It would be interesting to see if this link can be established experimentally. Moreover, by minimising the free energy with respect to both the order parameter field and the size and shape of the unit cell it is possible to predict the electrostriction of the blue phases, obtaining good qualitative and quantitative agreement with experiments. 

The major drawback of the Landau-de Gennes expansion is in the difficulty of knowing values of the expansion coefficients for any particular compound. Inevitably this programme becomes increasingly difficult as additional terms are added. Therefore the value of the theory is primarily in suggesting trends, and in identifying regions of parameter space where novel behaviour might be observed.

Although the major features of the equilibrium behaviour of the blue phases are understood, much less is known about their hydrodynamics. The current simulations are on tiny samples and it is necessary to guess suitable boundary conditions on the director field. However, the reponse of the blue phases  and their disclination networks to an imposed flow is fascinating, and a programme of developing the numerical approach in tandem with experiment is likely to uncover novel physics. 

Solving the equations of motion of the blue phases also makes it possible to investigate their dynamics under changes in an applied electric field. Understanding this is an important step to assisting device design. Moreover it is possible to follow the kinetics of transitions between the blue phases, as an external parameter such as the electric field is varied. This allows prediction of possible sequences of phases that will, in general, depend on the values of the free energy, whether there is a convenient path through phase space allowing any given phase to be accessed, and the speed with which the perturbation is applied. For the examples described here the structural evolution depended primarily on relaxation kinetics, with hydrodynamics playing a minor role. However, this may not always be the case.

Finally we return to blue phases that have been stabilised over a larger temperature range. One approach has been to use bimesogenic molecules, and there remain interesting questions as to whether the resulting blue phase is stable or metastable and in identifying the physical mechanisms behind the increased temperature range. The polymer stabilised blue phases can be interpreted in terms of the polymers pinning the disclinations. As numerical work on ever more complex fluids develops, there are many exciting questions about blue phases in mixtures of liquid crystals and polymers, colloids and nanoparticles, that will become increasingly accessible.

\acknowledgements{The authors would like to thank Davide Marenduzzo, Alexandre Dupuis and Enzo Orlandini for their help, advice and contributions to the research presented here. GPA would also like to thank Randall Kamien for discussions and acknowledges partial support from NSF Grant DMR05-47320.}



\clearpage

\clearpage

\clearpage

\clearpage

\clearpage

\clearpage

\clearpage

\clearpage

\clearpage

\end{document}